\documentclass[12pt,preprint]{aastex}
\usepackage{amssymb}
\usepackage{color}

\usepackage{graphicx}
\usepackage{natbib}

\usepackage{amssymb}
\usepackage{epstopdf}

\slugcomment{Not to appear in Nonlearned J., 45.}

\shorttitle{ Dependence of Hard X-ray Index on Eddington Ratio}
\shortauthors{Erlin Qiao \& B. F. Liu}

\begin{document}


\title{A Model for the Correlation of Hard X-ray Index with Eddington
Ratio in Black Hole X-ray Binaries}


\author{Erlin Qiao\altaffilmark{1} and B. F. Liu \altaffilmark{1}}
\affil{National Astronomical Observatories, Chinese Academy of
Sciences, Beijing 100012, China}

\email{qiaoel@nao.cas.cn}

\altaffiltext{1}{National Astronomical Observatories, Chinese
Academy of Sciences, Beijing 100012, China}

\begin{abstract}
Observations show that there is a positive correlation between
Eddington ratio $\lambda$ and hard X-ray index $\Gamma$ for $\lambda
\gtrsim 0.01$, and there is an anti-correlation between $\lambda$
and $\Gamma$ for $\lambda \lesssim 0.01$ in black hole X-ray
binaries (with $\lambda=L_{\rm bol}/L_{\rm Edd}$). In this work, we
theoretically investigate the correlation between $\Gamma$ and
$\lambda$ within the framework of disk-corona model. We improve the
model by taking into account all cooling processes including
synchrotron and self-Compton radiations in the corona,
Comptonization of the soft photons from the underlying accretion
disk, and the Bremsstrahlung radiations. Presuming that the coronal
flow above the disk can reach up to 0.1 Eddington rate at the outer
region,  we calculate the structure of the two-phase accretion flows
and the emergent spectra for accretion rates from 0.003 to 0.1. It
is found that at accretion rates larger than $\backsim 0.01$
Eddington rate a fraction of coronal gas condenses into the disk and
an inner disk can be sustained by condensation. In this case, the
X-ray emission is dominated by the scattering the soft photon from
the underlying disk in the corona. The emission from the inner disk
and corona can produce the positive correlation between $\lambda$
and $\Gamma$. While at accretion rates lower than $\backsim 0.01$
Eddington accretion rate, the inner disk vanishes completely by
evaporation, the accretion is dominated by ADAF, in which the X-ray
emission is produced by the Comptonization of the synchrotron and
bremsstrahlung photons of ADAF itself. The emission from ADAF can
produce the anti-correlation between $\lambda$ and $\Gamma$. We show
that our model can roughly explain the observed evolution of
$\Gamma_{\rm 3-25keV}$ with $L_{\rm 0.5-25 keV}/L_{\rm Edd}$ for the
black hole X-ray transient H1743-322 in the decay of 2003 from
thermal dominated state to low/hard state.

\end{abstract}



\section{Introduction}
A black hole X-ray binary (BHXBs) system is composed of a companion
star and a black hole which accretes gases from the companion star.
According to the spectral shape and   luminosity in X-ray band,
BHXBs are generally classified as five different spectral states. In
order of increasing luminosity these are quiescent state, low/hard
state, intermediate state, high/soft state and very high state (Esin
et al. 1997; 1998). Typically, the high/soft state is dominated by a
disk component around 1 $\rm keV$ and a weaker power-law tail with a
softer photon index $\Gamma \sim 2.5$. The low/hard state is
characterized by a power-law with a photon index $\Gamma \sim
1.5-2.1$ in the range of $2-10 \ \rm keV$ and an exponential cutoff
around 100 $ \rm keV$ (see review by Remillard \& McClintock 2006;
Done et al. 2007; and the references therein). The intermediate
state which is connected with the transition between the high/soft
state and low/hard state   exhibits complex spectral features
(Ebisawa et al 1994; Belloni et al. 1996). The observed spectrum of
this state is intermediate between high/soft state and low/hard
state. With the advancement of the observational technique, the
study on the quiescent state or the so-called 'off state' with flux
levels several order magnitude lower than that of the low/hard state
becomes possible. The spectrum of quiescent state is clearly
non-thermal, with a photon index a little softer than in the
low/hard state (Esin et al. 1997, and the references therein). At
the very high state, both the black-body component and the
non-thermal tail   become comparable (Van der Klis 1994; Gilfanov et
al. 1993). The different spectral states imply the different
accretion model, and the transition of the spectral state may mean
the change of the accretion patterns. It is proposed that the
different spectral states are primarily governed by the mass
accretion rate (Esin et al. 1997, 1998; Gilfanov et al. 2010; Liu et
al. 1999, 2002; Qiao \& Liu 2009, 2010, 2011; Maccarone 2003;
Meyer-Hofmeister et al. 2012). Homan et al. (2001) analyzed the data
of the black hole X-ray transient XTE J1550-564, taken with the {\em
Rossi X-ray Timing Explorer} between 1998 November 22 and 1999 May
20, during which the source went through different spectral state.
It's found that at least two physical parameters are necessary to
explain the behavior of XTE J1550-564. One of the parameter is
probably the mass accretion rate, and the other parameter is very
likely the (relative) size of a Comptonization region (Homan et al.
2001), which is thought to be associated with the strength of
relativistic jets (Fender et al. 2004, and the references therein).

Generally,  the accretion   at the high/soft state is dominantly via
a standard thin disk extending to the innermost stable circular
orbit (ISCO) with a weak corona above the disk (Pringle \& Rees
1972; Shakura \& Sunyaev 1973; Mitsuda et al. 1984; Frank et el.
2002); The ISCO is at $3R_{\rm S}$ (where $R_ {\rm S}=2GM/c^2$ is
the Schwarzschild radius, $G$ is gravitational constant, $c$ is
light speed, and $M$ is the central black hole mass). At the
low/hard state,   it is commonly believed that the accretion flows
are geometrically thick, optically thin, hot advection dominated
accretion flows (ADAF) within the region of a few  ten or hundred
Schwarzschild  radii from the central black hole + outer
geometrically thin, optically thick, cool accretion disk (Esin et
al. 2001; McClintock et al. 2001). The accretion flow can account
for the observed UV and X-ray emission, but it substantially
underpredicts the radio emission. The accretion-jet model was
proposed to fit the multi-wavelength observations of the black hole
X-ray binary XTE J1118+480 (Yuan et al. 2005, 2007). The jet model
was also proposed for the low/hard state, in which both the radio
and X-ray emission originate from the synchrotron emission of the
jet (Markoff et al. 2001). For the observed intermediate state, the
accretion pattern is still not clear. The iron $K_{\alpha}$ line,
the reflection component and also the timing property might imply
that the cool gas resides in the innermost region around the black
hole (Meyer et al. 2007). At the quiescent state, with much lower
accretion rate, the accretion flows have two distinctive zones (1) a
standard thin disk at a radius larger than 3000 $R_{\rm S}$   for
the optical and UV radiation  (2)   an inner ADAF   for the X-ray
emission (Narayan et al. 1996). Recently, observations show that
there is a positive correlation between Eddington ratio $\lambda$
and hard X-ray index $\Gamma$ for   $\lambda>0.01$ (Sobolewska et
al. 2011; Kalemci et al. 2006), and there is an anti-correlation
between $\lambda$ and $\Gamma$ for
 $\lambda \lesssim 0.01$ ($\lambda=L_{\rm bol}/L_{\rm Edd}$,
with $L_{\rm bol}$ being the bolometric luminosity, $L_{\rm
Edd}=1.26 \times 10^{38} m \ \rm erg \ s^{-1}$ being the Eddington
luminosity, $m$ being the central black hole mass scaled with solar
mass $M_{\odot}$) (Wu \& Gu 2008). It's interesting that this
correlation is also found in active galactic nuclei, i.e., when
$\lambda \gtrsim 0.01$, the    hard X-ray spectrum  softens with
Eddington ratio (Porquet et al. 2004; Shemmer et al. 2006; Saez et
al. 2008; Sobolewska \& Papadakis 2009; Cao 2009; Liu et al. 2012;
Zhou et al. 2010; Veledina et al. 2011), but when $\lambda \lesssim
0.01$ the hard X-ray spectrum hardens with the Eddington ratio
(Constantin et al. 2009; Gu \& Cao 2009; Younes et al. 2011, 2012;
Xu 2011). For the Comptonization models, it's argued that the above
correlations mean the  ratio of  Compton luminosity to seed photon
luminosity, $\ell_{h} / \ell_{s}$, changes with Eddington ratio,
which is consistent with the scenario that the seed photons    are
from the synchrotron radiation at lower Eddington ratio and from the
truncated thin disk at higher Eddington ratio (Sobolewska et al.
2011). Physically, the change   from negative to positive
 correlations may imply   a change of the accretion patterns
(Esin et al. 1997, 1998; Yuan et al. 2007).

In this work, by considering the vertical structure of the ADAF, a
disk evaporation/ condensation model is proposed to give     a
unified description to the two different correlations in the black
hole X-ray binaries. We presume that the coronal flow above the disk
can reach up to 0.1 Eddington accretion rate at the outer region. It
is found that at mass accretion rates larger than $\backsim 0.01$ a
fraction of coronal gas condenses into the disk and an inner disk
can be sustained by condensation rather than be swallowed by the
black hole within the viscous time scale (Liu et al. 2007; Taam et
al. 2008; Liu et al. 2011). The size of the inner remnant disk
$r_{\rm d}$ is governed by the accretion rate. The geometry of the
accretion flows are described as an inner disk and corona + outer
ADAF. During this phase, a positive correlation between $\lambda$
and $\Gamma$ will be predicted. Compared with Meyer et al. (2006),
Liu et al. (2006, 2007), Taam et al. (2008) and Qiao \& Liu (2012),
in this work, we self-consistently consider the cooling of the soft
photons from the underlying accretion disk to the corona, and the
bremsstrahlung cooling, synchrotron cooling and the corresponding
self-Compton cooling of the corona itself. With decrease of the mass
accretion rate, the size of the inner disk decreases, eventually the
inner disk vanishes completely. The X-ray emission is dominated by
the inner ADAF. In the ADAF model, X-ray emission is produced by
Compton scattering of the bremsstrahlung and synchrotron photons of
the ADAF itself. During this phase, an anti-correlation between
$\lambda$ and $\Gamma$ will be predicted (Esin et al. 1997, 1998).

In a word, we propose a self-consistent model which can
quantitatively describe both the positive correlation in higher
Eddington ratio and the anti-correlation in lower Eddington ratio in
the black hole X-ray binaries. In section 2, we introduce the model.
The numerical results are presented in section 3. Some Comparisons
with observations are shown in section 4.  Section 5 is the
conclusion.

\section{The model}
\subsection{Formation of the inner disk and corona}
\label{condensation} Recent investigations (Liu et al. 2006, 2007;
Meyer et al. 2007) reveal that when the accretion rate is not far
below the maximal evaporation rate an inner disk separated from the
outer disk by a coronal region could also exist, leading to an
intermediate state of black hole accretion. The onset of an
intermediate state occurs as the accretion rate decreases just below
the maximal evaporation rate. At this time, disk truncation by
evaporation sets in near the region where the evaporation rate is
maximal. A coronal gap appears and widens with a further decrease in
the accretion rate with the inner cool disk reduced in extent.
Because of diffusion, the inner disk can not survive for a time
longer than a viscous time (which is only a few days in the inner
disk   of BHXBs) unless matter continuously condenses from
the ADAF onto the cool inner disk. In the following we investigate
the interaction between the disk and the corona/ADAF, showing the
conditions under which matter condenses onto the inner disk, thereby
maintaining a cool disk in the inner region.

For simplicity, we consider a hot ADAF-like corona described by the
self-similar solution of Narayan \& Yi (1995b) above a geometrically
thin standard disk around a central black hole. In the corona,
viscous dissipation leads to ion heating, which is partially
transferred to the electrons by means of Coulomb collisions. This
energy is partially radiative away by bremsstrahlung, synchrotron
emission, the    self-Compton emission and the Compton scattering of
the soft photons from the underling disk, and partially transferred
down to the transition layer between the disk and the corona. If the
density in this transition layer is too low to efficiently radiate
the energy, cool matter is heated up and evaporated into the corona.
On the other hand, if the density in this layer is sufficiently
high, the conductive flux is completely radiated away, the coronal
mass is   over-cooled, and condenses   partially to the disk.   The
evaporation or the condensation goes on until  an equilibrium
density
  is established. The gas in the corona still retains angular
momentum and with the role of viscosity will differentially rotate
around the central object. By friction the gas looses angular
momentum and drifts inward thus continuously drains mass from the
corona towards the central object.  Therefore, mass is accreted to
the central object partially through the corona and partially
through the disk. The  pressure $p$, the electron number density
$n_{\rm e}$, viscous heating rate $q^{+}$, and the sound speed of
ADAF $c_{\rm s}$ depend on black hole mass $m$, mass accretion rate
$\dot m_{\rm c}$, the viscosity parameter $\alpha$, the  parameter
  describing the magnetic fields in the accretion flows,
$\beta$ (defined as the ratio of gas pressure to the total
pressure), and the distance from the black hole $r$, in the form
(Narayan \& Yi 1995b),
\begin{eqnarray}\label{para}
\begin{array}{l}
p=1.71\times10^{16}\alpha^{-1}c_{1}^{-1}c_{3}^{1/2}m^{-1}\dot m_{\rm c} r^{-5/2} \ \ \ \rm g \ cm^{-1} \ s^{-2}, \nonumber \\
n_e=2.00\times10^{19}\alpha^{-1}c_1^{-1}c_{3}^{-1/2}m^{-1} \dot m_{\rm c} r^{-3/2}\ \ \ \rm cm^{-3},  \nonumber \\
q^{+}=1.84\times 10^{21}\varepsilon^{'}c_{3}^{1/2}m^{-2}\dot
m_{\rm c} r^{-4}\ \ \rm ergs \ cm^{-3} \ s^{-1}  \nonumber, \\
c_s^2 =4.50\times10^{20}c_{3}r^{-1} \ \ \rm cm^{2} \ s^{-2}   ,
%\nonumber
\end{array}
\end{eqnarray}
where   $\dot m_{\rm c}$ is the coronal/ADAF mass accretion rate scaled with
Eddington accretion rate $\dot M_{\rm Edd}$ ($\dot M_{\rm
Edd}$=$1.39\times 10^{18}m$), $r$ is the radius scaled with
Schwarzschild radius $R_{\rm S}$, and
\begin{equation}\label{coef}
\begin{array}{l}
{c_1}={(5+2\varepsilon^{'}) \over {3\alpha^2}}g(\alpha,\varepsilon^{'}),\\
\\
{c_3}={2\varepsilon(5+2\varepsilon^{'})\over {9\alpha^2} } g(\alpha,\varepsilon^{'}),\\
\\
{\varepsilon{'}}={\varepsilon\over f}={1\over f} \biggl({{5/3-\gamma}\over {\gamma-1}}\biggr),\\
\\
g(\alpha,\varepsilon^{'})=\biggl[ {1+{18\alpha^2\over (5+2\varepsilon^{'})^{2}}\biggr]^{1/2}-1},\\
\\
\gamma={{32-24\beta-3\beta^2}\over {24-21\beta}},
\end{array}
\end{equation}
with $f$ the advection fraction. The heat flux transferred from the
typical corona/ADAF to the transition layer is derived from energy
balance in the corona, i.e.,
\begin{eqnarray}\label{energy}
\Delta F_{\rm c}/H=q_{\rm ie} -q_{\rm rad},
\end{eqnarray}
where $\Delta F_{\rm c}$ refers to the flux transferred from the
upper boundary of the corona to the interface of the transition
layer. $H$ is the scaled height   of the corona, given approximately by
$H=(2.5c_{3})^{1/2}rR_{\rm S}$. $q_{\rm ie}$ is the energy transfer
from ions to electrons (Stepney 1983), and can be approximately
expressed for two-temperature advection-dominated
hot flow as   (Liu et al. 2002),
\begin{eqnarray}\label{qie}
q_{\rm ie} & = & (3.59\times 10^{-32} {\rm g\ cm^5\ s^{-3}\ K^{-1}})
n_e n_i T_i {\left(\frac{k T_e}{m_e c^2}\right)}^{-3/2},
%       & = & 1.05\times 10^{35}
%T_e^{-3/2} \alpha^{-2}m^{-2} \dot m^{2}r^{-4} {\, \rm{ g
%cm^{-1}s^{-3}deg^{3/2}}},
\end{eqnarray}
%$\delta$ refers to the fraction of direct heating to the electrons in the corona,
$q_{\rm rad}(n_{\rm e}, T_{\rm e})=$ $q_{\rm
brem}+q_{\rm syn}+q_{\rm cmp}+q_{\rm excmp}$ is the radiative
cooling rate of the corona. With $q_{\rm brem}$, $q_{\rm syn}$ and
$q_{\rm Cmp}$  the bremsstrahlung cooling rate, synchrotron
cooling rate and the corresponding self-Compton cooling rate
respectively. $q_{\rm brem}$, $q_{\rm syn}$ and $q_{\rm Cmp}$ are
all the functions of   electron number  density $n_{\rm e}$ and
electron temperature $T_{\rm e}$ (Narayan et al. 1995b; Manmoto et
al, 1997). $q_{\rm excmp}$ is the Compton cooling rate of the
underling disk   photons to the corona, given by
\begin{eqnarray}\label{cmp}
q_{\rm excmp}={4kT_{\rm e}\over {m_{\rm e}c^2}}n_{\rm
e}\sigma_{T}cu,
\end{eqnarray}
with $u$ the soft-photon energy density from the underling thin
disk. $u$ is expressed in terms of the effective temperature in the
local disk as $u={a T_{\rm eff}^{4}(r)}={a} \bigg \lbrace 2.05T_{\rm
eff,max} \bigg({3 \over r} \bigg)^{3/4} \bigg[1-\bigg({3 \over r}
\bigg)^{1/2} \bigg]^{1/4} \bigg \rbrace^{4} $ (where $a$ is
radiation constant). The conductive flux is given by the formula
$F_{\rm c}(z)=k_{0}T_{\rm e}^{5/2}dT_{\rm e}/dz$ (Spitzer 1962). For
simplicity, we assume that the conductive flux transferred from the
upper boundary of the corona to the interface of the transition
layer is as follows,
\begin{eqnarray}\label{spizter}
\Delta F_{\rm c} & = & k_{0}T_{\rm em}^{5/2}(T_{\rm em}-T_{\rm
cpl})/H  \nonumber \\
                 & = & k_{0}T_{\rm em}^{7/2}(1- T_{\rm cpl}/T_{\rm
                 em})/H  \nonumber \\
                 & \backsimeq & k_{0}T_{\rm em}^{7/2}/H,
\end{eqnarray}
where $T_{\rm em}$ is the temperature of the electron at upper
boundary and refers to the maximum temperature within a given column
at a distance $r$. $T_{\rm cpl}$ is the coupling temperature at the
transition layer, and is determined by assuming viscous heating rate
$q^{+}$ and compressive heating rate $q^{\rm c}$ balanced with the
transfer of heat rate from the ions to the electrons $q_{\rm ie}$,
i.e., $q_{\rm ie}=q^{+}+q^{\rm c}$, where $q^{c}={1 \over
(1-\beta)}q^{+}$ (Esin 1997; Meyer et al. 2007). The eq.
(\ref{spizter}) approximately holds for $T_{\rm cpl}/T_{\rm em}$
much less than one (Meyer et al. 2006; Meyer et al. 2007; Liu et al.
2007). As the temperature of electron at a given distance along
vertical direction in the main body of the corona is approximately
constant (Liu et al. 2002), in eqs. (\ref{qie}) and (\ref{cmp}) the
local temperature of the electron $T_{\rm e}$ is replaced by $T_{\rm
em}$ throughout our calculations. Given the central black hole mass
$m$, mass accretion rate in the corona $\dot m_{\rm c}$, the value
of viscosity parameter $\alpha$, magnetic parameter $\beta$, and the
maximum effective temperature of the accretion disk $T_{\rm
eff,max}$, by combing equations (\ref{para}), (\ref{energy}),
(\ref{qie}), (\ref{cmp}) and (\ref{spizter}), the temperature $
T_{\rm em}$ and the heat flux transferred from the typical
corona/ADAF to the transition layer $\Delta F_{\rm c}$ are solved.
If the flux at the maximum height of the corona/ADAF is assumed to
be approximately zero, the flux arriving at the interface of the
transition layer $F_{\rm c}^{\rm ADAF}$ is $\Delta F_{\rm c}$.

Following the work of Liu et al. (2007), the energy balance in the
transition layer is determined by the incoming conductive flux,
bremsstrahlung radiation flux, and the enthalpy flux carried by the
mass condensation flow,
\begin{equation}\label{energy-layer}
\frac{d}{dz} \left[\dot m_z \frac{\gamma}{\gamma-1}
\frac{1}{\beta}\frac{\Re T}{\mu} + F_c \right] = -n_e n_i
\Lambda(T).
\end{equation}

This determines the condensation rate per unit area, which is given
by (Meyer et al. 2007; Liu et al. 2007),
\begin{equation}\label{cnd-general}
\dot m_z= {{\gamma-1} \over \gamma}\beta {{-F_{c}^{ADAF}} \over {\Re
T_{i}/ \mu_{i}}}(1-\sqrt{C}),
\end{equation}
with
\begin{equation}\label{C}
C \equiv\kappa{_0} b \left(\frac{0.25\beta^2 p_0^2}{k^2}\right)
\left(\frac{T_{\rm {cpl}}}{F_c^{\rm{ADAF}}}\right)^2.
\end{equation}

From eq. (\ref{C}), for $C=1$, a critical radius $r_{\rm d}$ is
derived, from which inwards part of the corona/ADAF matter condenses
onto the disk, and a disk-corona system forms. $r_{\rm d}$ is called
as condensation radius in this paper. Combing eqs.
(\ref{cnd-general}) and (\ref{C}), the integrated condensation rate   in unit of
 Eddington rate from $r_{\rm d}$ to any radius $r$ of the
disk reads,
\begin{equation}\label{condensation}
\dot m_{\rm cnd}(R)= \int_{R_i}^{R_d} {4\pi R \over \dot M_{\rm
Edd}} \dot m_z dR.
\end{equation}
  According to mass conservation, if the initial mass accretion rate
in the corona is $\dot m$ ($\dot m$ = $\dot m_{\rm c} (r_{\rm d})$),
the mass accretion rate in the corona is a function of distance,
\begin{equation}\label{mdot-corona}
\dot m_{\rm c}(R)=\dot m -\dot m_{\rm cnd}(R).
\end{equation}
The luminosity of the corona are derived by integrating the corona
region,
\begin{equation}\label{Luminosity}
L_{\rm c, in}=\int_{\rm 3 R_{\rm S}}^{R_{\rm d}} q_{\rm rad}
\bigg(\dot m_{\rm c}(R), R)\bigg) H 4\pi R dR.
\end{equation}

If the irradiation of the corona to the disk is considered, the
effective temperature of the accretion disk including both the
accretion fed by the condensation and the irradiation of the corona
is given by the following formula (see Liu et al. 2011; Qiao \& Liu
2012 for details),
\begin{eqnarray}\label{trp}
T_{\rm eff}(r)= 2.05T_{\rm eff,max}^{\prime} \bigg({3 \over r}
\bigg)^{3/4}
\bigg[1-\bigg({3 \over r} \bigg)^{1/2} \bigg]^{1/4} \nonumber \\
\times \bigg[{1+6L_{\rm c,in}(1-a) \over \dot M_{\rm cnd}c^2}
{H_{\rm s} \over 3R_{\rm s}}\bigg]^{1/4} \nonumber \\
=2.05T_{\rm eff,max} \bigg({3 \over r} \bigg)^{3/4} \bigg[1-\bigg({3
\over r} \bigg)^{1/2} \bigg]^{1/4},
\end{eqnarray}
where $a$ is albedo which is defined as the energy ratio of
reflected radiation from the surface of the thin disk to incident
radiation upon it from the corona, the expression of $T_{\rm
eff,max}$   is as follows,
\begin{eqnarray}\label{tmaxp}
T_{\rm eff,max}=T_{\rm eff,max}^{\prime} \bigg[{1+6L_{\rm c,in}(1-a)
\over \dot M_{\rm cnd}c^2} {H_{\rm s} \over 3R_{\rm s}}\bigg]^{1/4}.
\end{eqnarray}
$T_{\rm eff,max}^{\prime}$ refers to the maximum effective
temperature from disk accretion which is reached at $r_{\rm
tmax}=(49/12)$. The expression of $T_{\rm eff,max}^{\prime}$ is
given as (Liu et al. 2007),
\begin{eqnarray}\label{tmax}
T_{\rm eff, max}^{\prime}=0.2046 \bigg({m\over 10}
\bigg)^{-1/4}\bigg[{\dot m_{\rm cnd(r_{\rm tmax})}\over
0.01}\bigg]^{1/4} \ \rm keV.
\end{eqnarray}

Given the black hole mass $m$, the initial mass accretion rate in
the corona $\dot m$ ($\dot m$ = $\dot m_{\rm c} (r_{\rm d})$), the
viscosity parameter $\alpha$, magnetic parameter $\beta$, and albedo
$a$, a maximum effective temperature of the accretion disk $T_{\rm
eff,max}$ is assumed to calculate the condensation rate from eq.
(\ref{condensation}) and corona luminosity from eq.
(\ref{Luminosity}), with which a new effective temperature $T_{\rm
eff,max}$ is derived from eqs. (\ref{tmaxp}) and (\ref{tmax}). An
iteration is made till the presumed temperature is consistent with
the derived value, so we find a self-consistent solution of the
inner disk and corona, including the size of the inner disk $r_{\rm
d}$, the condensation rate $\dot m_{\rm cnd}(r)$, the temperature of
the corona $T_{\rm em}(r)$ and Compton scattering optical depth
$\tau_{\rm es}(r)$ in radial direction. With the derived structure
of the inner disk and corona, the emergent spectrum of the inner
disk-corona system is calculated out (Qiao \& Liu 2012).

\subsection{The ADAF model}
Outside the condensation radius $r_{\rm d}$ of inner disk, the
accretion flows are supposed to be in the form of advection
dominated accretion flows (ADAF) or radiatively inefficient
accretion flow (RIAF) (Rees et al. 1982; Narayan \& Yi 1994; Narayan
\& McClintock 2008, and the references therein). The self-similar
solution of ADAF is first proposed by (Narayan \& Yi 1994; Narayan
\& Yi 1995b), with which the spectra of transient source A0620-00
and V404 Cyg with lower luminosity are well fitted (Narayan et al.
1996). Later, the global solution of ADAF are conducted by several
authors (Narayan et al. 1997; Manmoto 1997, 2000; Yuan et al. 1999,
2000). All of them show that the self-similar solution is a good
approximation at a radius far enough from the ISCO. For simplicity,
in the present paper, the self-similar solution of ADAF, as in eq.
(\ref{para}), is adopted (Narayan \& Yi 1995a,b). The structure of
the ADAF can be derived by specifying the parameter $m$, $\dot m$
($\dot m=\dot m_{\rm c} (r_{\rm d})$), $\alpha$, and $\beta$. All
the radiative processes, including bremsstrahlung cooling $q_{\rm
brem}$, synchrotron cooling $q_{\rm syn}$, and the corresponding
self-Compton cooling  $q_{\rm cmp}$ are considered in the
calculation of the self-similar solution. With the derived
structure, emergent spectrum of ADAF is calculated (Qiao \& Liu
2010).

Combing the contribution of the inner disk-corona system and the
outer ADAF, we get the total spectrum of the black hole accreting
system. For the inner disk-corona   region, we calculate both
scattering of the soft photons form the underlying disk, and scattering of the
synchrotron and bremsstrahlung photons
from the corona itself.   Because we
focus on the hard X-ray emission, throughout this   study, we neglect
the contribution of outer truncated disk to the total emergent
spectrum.

\section{Numerical results}
\begin{figure*}
\includegraphics[width=85mm,height=70mm,angle=0.0]{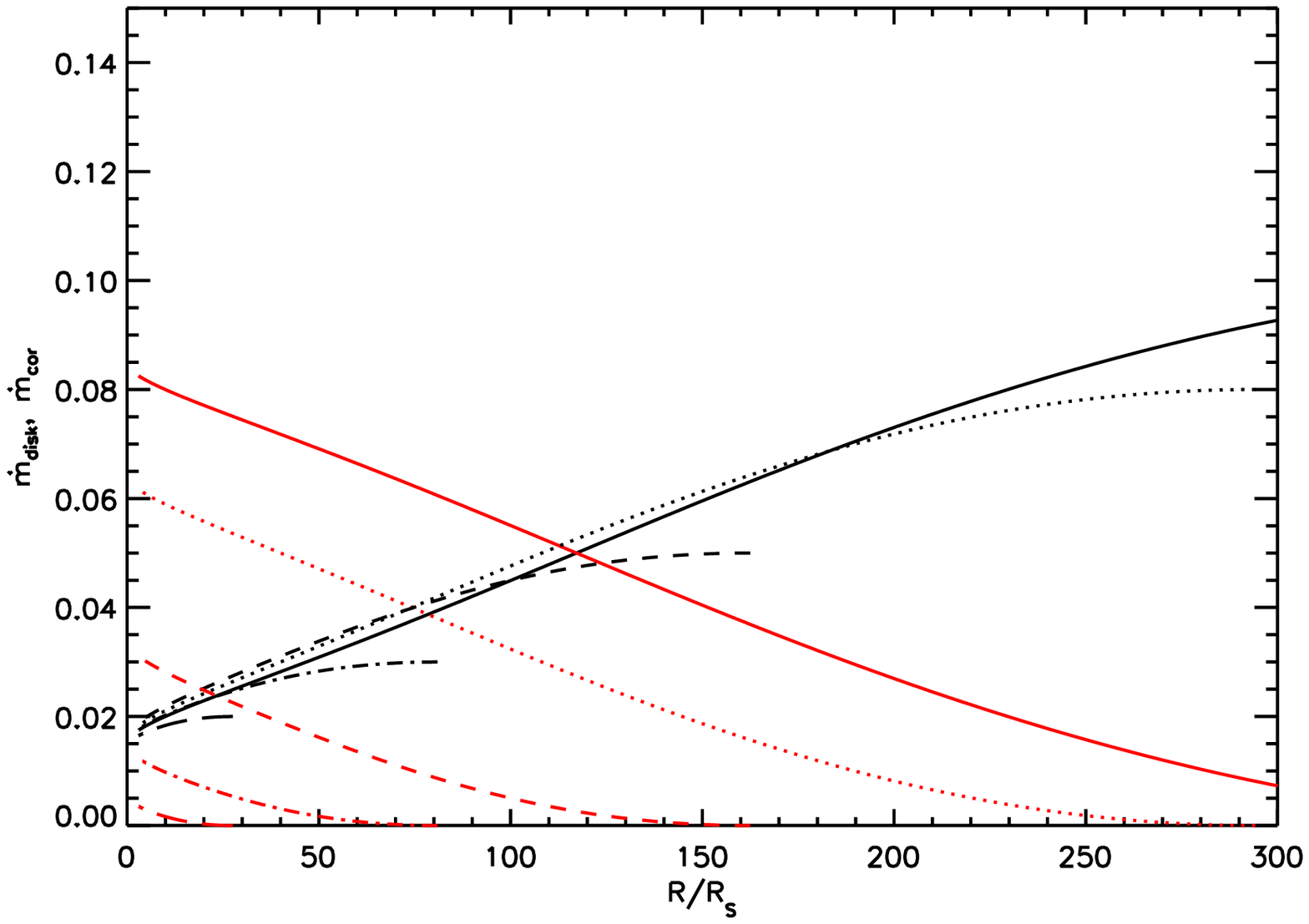}
\caption{\label{mdotcorona} Mass accretion rate in the accretion
disk and the mass accretion rate in the corona as functions of
radius. The red line and black line represent the mass accretion
rate in the disk and the mass accretion rate in the corona
respectively. The solid line, dotted line, dashed line,
dotted-dashed line and long-dashed line are for $\dot m =0.1$,
$0.08$, $0.05$, $0.03$ and $0.02$ respectively. In the calculation,
$m=10$, $\alpha=0.3$, $\beta=0.8$ and $a=0.15$ are adopted.}
\end{figure*}

In  the calculation, the  black hole mass $m=10$ and $\alpha=0.3$
are adopted. We fix $\beta=0.8$ as suggested by the simulations of
turbulence driven by the magneto-rotational instability in a
collisionless plasma (Sharma et al. 2006; Meyer et al. 2007). The
albedo is often very low, $a \sim 0.1-0.2$ (e.g., Magdziarz \&
Zdziarski 1995; Zdziarski et al. 1999), which means most of the
incident photons from the corona are absorbed by the accretion disk,
and reradiated as blackbody radiation. The effect of albedo $a$ to
the structure of disk and corona, and the corresponding spectra
was studied by Qiao \& Liu (2012). Here, $a=0.15$ is adopted. It's
found that the accretion flows are in the form of inner disk and
corona + outer ADAF if the mass accretion rates are in the range $
0.01 \lesssim \dot m \lesssim 0.1$. The inner disk vanishes   when
$\dot m \lesssim 0.01 $ and the accretion flows exist in the form of
ADAF. We list the numerical results in Table 1. In order to show the
mass distribution of the inner disk-corona system in the radial
direction, we plot the mass accretion rate in the accretion disk and
the mass accretion rate in the corona as a function of radius in
Figure \ref{mdotcorona}. It is clearly seen that, with increase of
the mass accretion rate, more gases in the corona   condense onto
the disk, and the accretion rate in the disk increases.

\begin{figure*}
\includegraphics[width=85mm,height=70mm,angle=0.0]{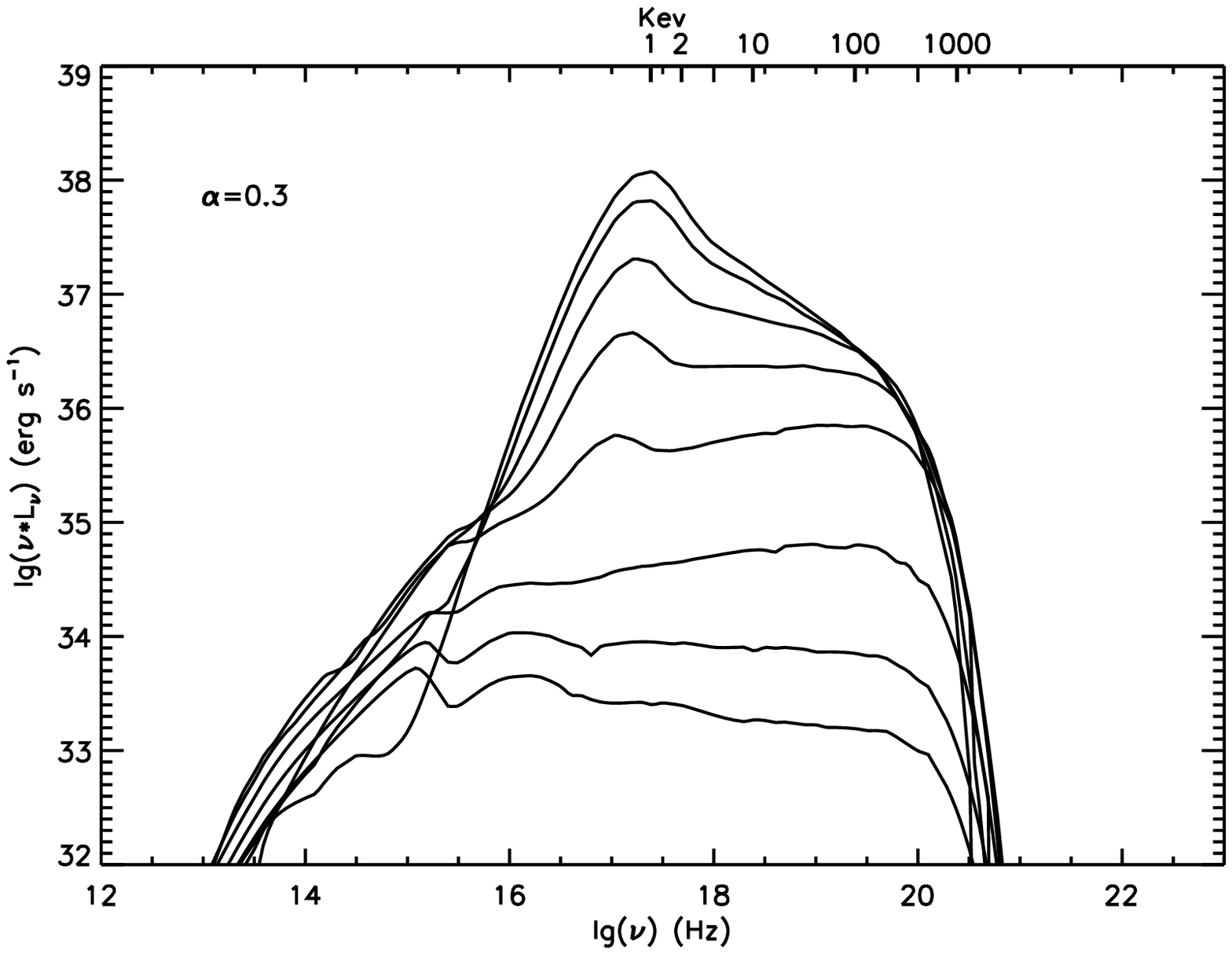}
\caption{\label{sp}  The emergent spectra for different mass
accretion rate. In the calculation, $m=10$, $\alpha=0.3$,
$\beta=0.8$ and $a=0.15$ are adopted. From the bottom up, the mass
accretion rates are $\dot m=0.003$, $0.005$, $0.01$, $0.02$, $0.03$,
$0.05$, $0.08$, $0.1$ respectively.}
\end{figure*}

\begin{figure*}
\includegraphics[width=85mm,height=70mm,angle=0.0]{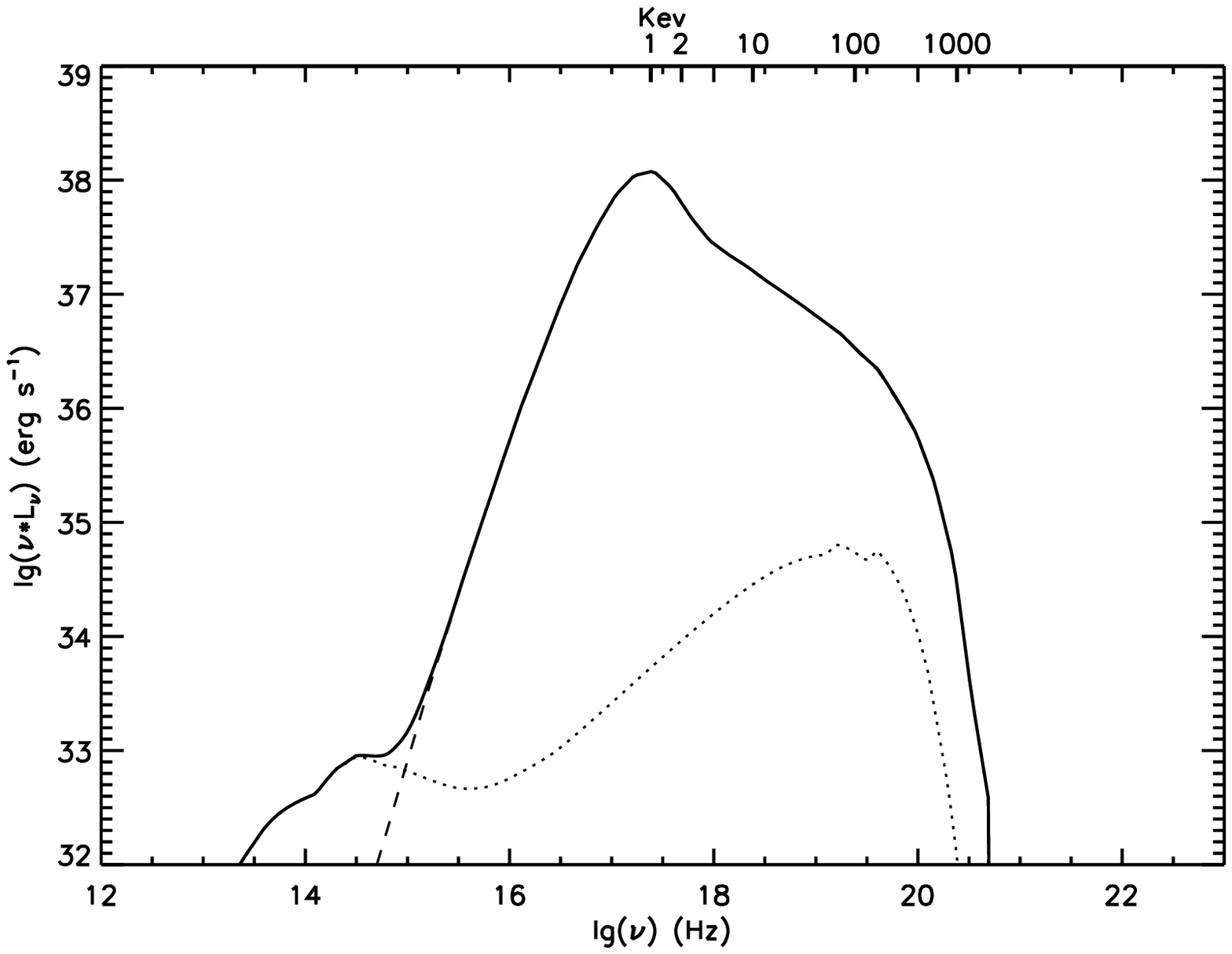}
\includegraphics[width=85mm,height=70mm,angle=0.0]{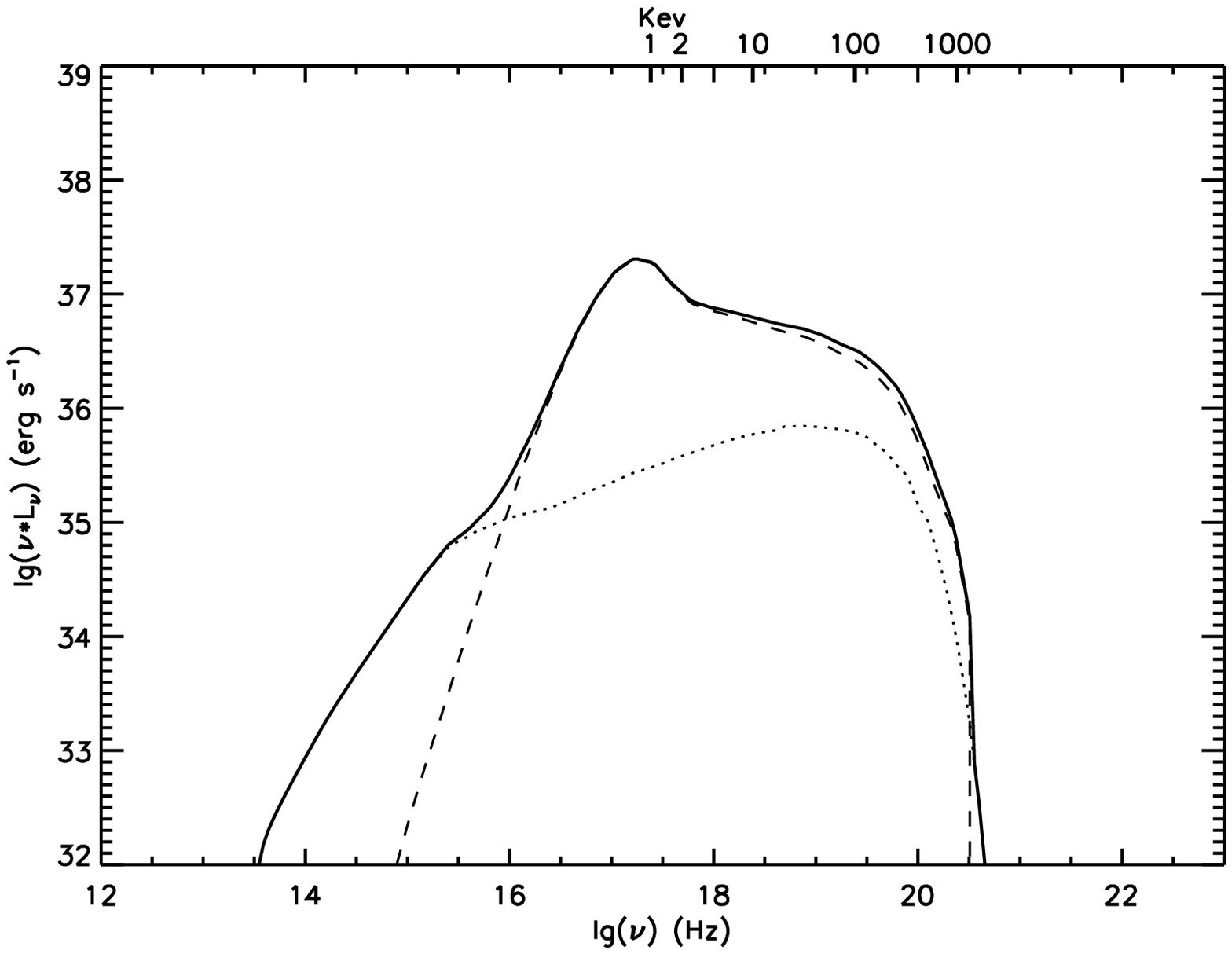}
\includegraphics[width=85mm,height=70mm,angle=0.0]{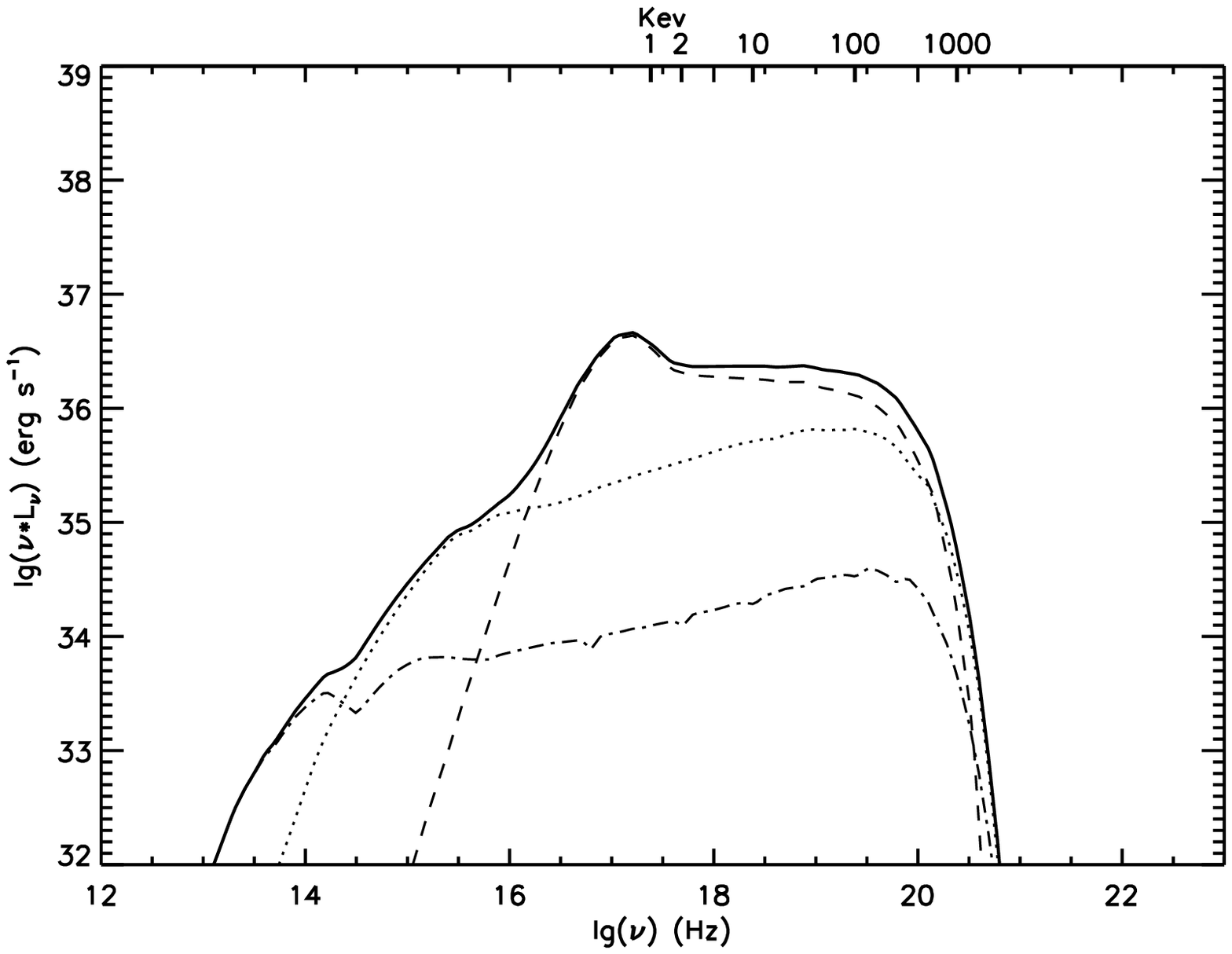}
\includegraphics[width=85mm,height=70mm,angle=0.0]{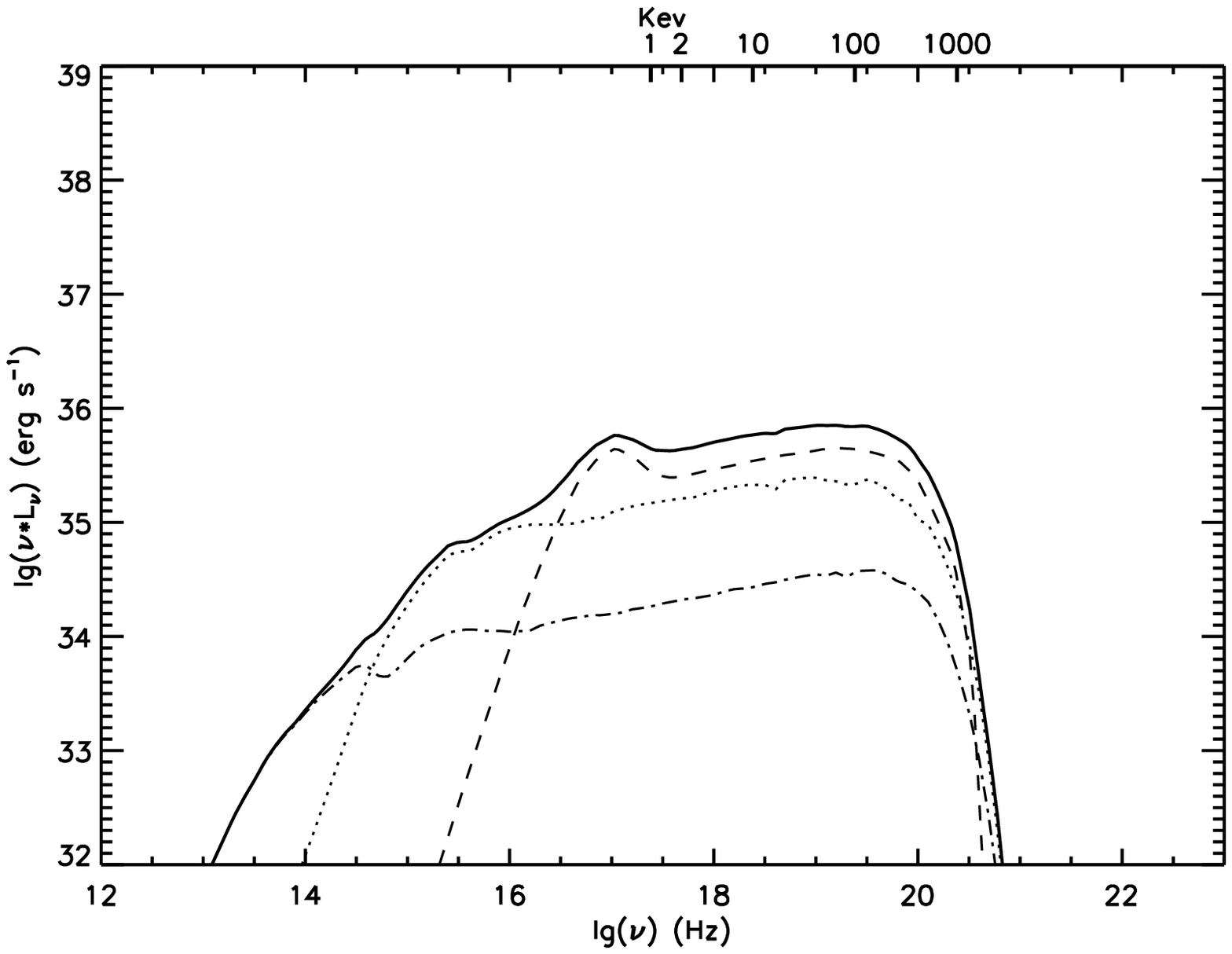}
\caption{\label{spmdot}  The emergent spectra for different mass
accretion rate. In all the plots, $m=10$, $\alpha=0.3$, $\beta=0.8$
and $a=0.15$ are adopted. The first plot is for $\dot m =0.1$. The
solid line is the total spectrum, the dashed line is the Compton
spectrum produced by scattering the soft photons from the underlying
disk in the corona, the dotted line is Compton spectrum produced by
scattering the synchrotron and bremsstrahlung photons of the corona
itself in the corona. The second plot is for $\dot m =0.05$. The
third and the forth plots are for $\dot m =0.03$ and $\dot m =0.02$
respectively. The dot-dashed line in the third and the forth plots
are the contribution of the ADAF beyond the condensation radius.}
\end{figure*}

The emergent spectra with mass accretion rate are plotted in Figure
\ref{sp}   for mass accretion rates  $\dot m=0.1$, $0.08$,
$0.05$, $0.03$, $0.02$, $0.01$, $0.005$, $0.003$. The contribution
of different components to the spectra for different mass accretion
rate   is shown in figure \ref{spmdot}. The first plot in
Figure \ref{spmdot} is for $\dot m=0.1$. The solid line is the total
spectrum, the dashed line is the Compton spectrum produced by
scattering the soft photon from the underlying disk in the corona,
the dotted line is Compton spectrum produced by scattering the
synchrotron and bremsstrahlung photons of corona itself in the
corona. It's   clear that the emergent spectrum is completely
dominated by the scattering the soft   photons from the
underlying disk in the corona for $\dot m=0.1$. This is because, for
high mass accretion rate, due to the very strong Comptom cooling,
most of the hot coronal gas condenses onto the underlying disk, the
density of soft photons from the disk   is much higher than
that of the synchrotron and bremsstrahlung photons from the corona.
In this case, the emission is dominated by the disk. With decrease
of the mass accretion rate to $\dot m=0.05$, the condensation rate
decreases, the relative strength of   Compton scattering  of
the soft photons from the underlying disk to   that from the
synchrotron and bremsstrahlung photons of corona itself  decreases,
   which can be seen from  the second plot in Figure
\ref{spmdot}.   With decrease of the mass accretion rate
further, the   relative strength of the corona decreases
continuously, as shown in the third and forth plots for $\dot
m=0.03$ and $\dot m=0.02$ as   examples. The dot-dashed line
in the third and the forth plots is the component from the outer
ADAF. Ultimately, when $\dot m \lesssim 0.01 $, the inner disk
  vanishes completely, and emission is dominated by the ADAF,
in which the soft photons are only from the synchrotron and
bremsstrahlung of the ADAF itself.

\begin{figure*}
\includegraphics[width=85mm,height=70mm,angle=0.0]{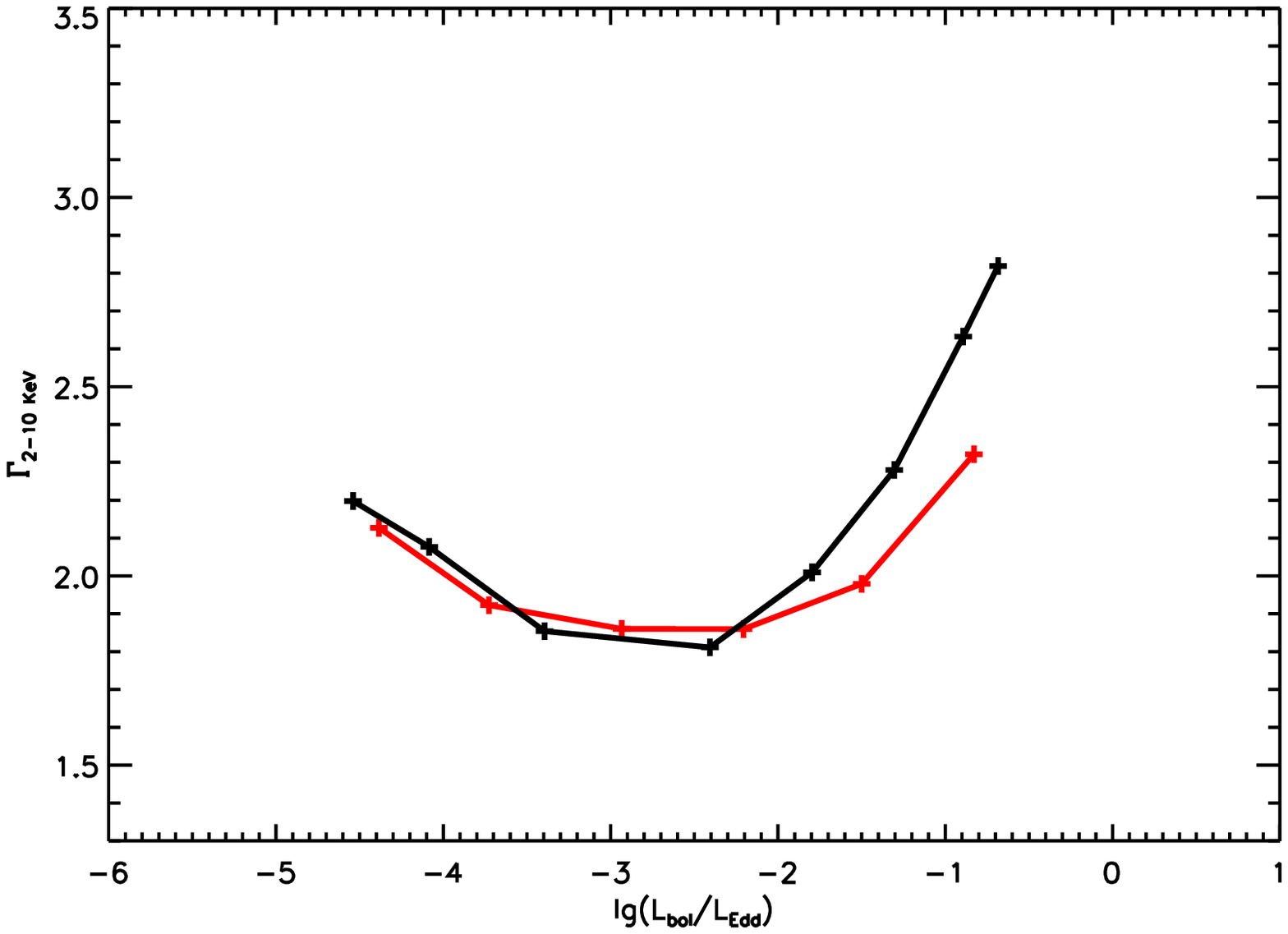}
\caption{\label{L-gama}} The dependence of hard X-ray photon index
$\Gamma_{\rm 2-10keV}$ on Eddington ratio $\lambda$. The black line
is for $\alpha=0.3$. The red line is for $\alpha=0.4$. All the
calculation, $m=10$, $\beta=0.8$ and $a=0.15$ are adopted.
\end{figure*}

From the emergent spectrum in Figure \ref{sp}, it's also found that,
with decrease of mass accretion rate   from 0.1 to 0.02, the
hard X-ray photon index $\Gamma_{\rm 2-10 keV}$ decreases from
  2.8 to 1.81.   Then $\Gamma_{\rm 2-10 keV}$
increases up to $2.20$ when $\dot m$ is down to $0.003$. The
evolution of  $\Gamma_{\rm 2-10 keV}$ with Eddington ratio $\lambda$
is plotted in figure \ref{L-gama} with black line. We can see that
there is a positive correlation between Eddington ratio $\lambda$
and hard X-ray index $\Gamma_{\rm 2-10 keV}$ in the range of
$\lambda \gtrsim 0.01$, and there is an anti-correlation between
$\lambda$ and $\Gamma$ for $\lambda \lesssim 0.01$.   The turning
point is at $\lambda \sim 0.01$ and $\Gamma_{\rm 2-10 keV} \sim
1.8$.  The above correlation can be understood as follows. At $\dot
m =0.1$, because of the very strong Compton cooling, most of the
coronal matter ($\sim 82\%$) condenses onto the disk. At this case,
the emission is dominated by the disk, so a softer X-ray spectrum is
predicted. With decrease of the mass accretion rate, both the size
of the inner disk and the amount of the coronal matter condensed to
the disk decrease. The decreased condensation rate means that the
relative strength of the disk luminosity to the coronal luminosity
decreases, which will make the spectrum  harder. Meanwhile, an
increased contribution of the outer ADAF to the spectrum will also
make the spectrum harder. So, during this phase, a positive
correlation between $\lambda$ and $\Gamma_{\rm 2-10 keV}$ is
predicted with decreasing of mass accretion rate.

Decreasing the mass accretion rate further, the size of the inner
disk decreases continually. Ultimately, the inner disk vanishes
completely   at some critical accretion rate, and the outer
disk   is suppressed   at a very large radius. The
X-ray emission is dominated by the inner ADAF.   In this
case, the Comptonization of the synchrotron and bremsstrahlung
photons of ADAF itself is the dominated cooling mechanism. With
decrease of the mass accretion rate,   the temperature of the
ADAF very weakly depends on the mass accretion rate, i.e. $T_{\rm e}
\propto \dot m^{-1/14}$ (Mahadevan 1997).    However, the
Thomson scattering optical depth goes down (Narayan \& Yi 1995b)
\begin{equation}\label{tau}
\tau_{\rm es}=12.4 \alpha^{-1} c_{1}^{-1} \dot m_{} r^{-1/2},
\end{equation}
where $\dot m_{}$ is mass accretion rate scaled with Eddington rate
and  $c_{1} \backsimeq 0.5$. This results in a decrease of the
Compton parameter $y$ with decrease of accretion rate. Consequently,
the hard X-ray index increases. So, during this phase, an
anti-correlation between $\lambda$ and $\Gamma_{\rm 2-10keV}$ is
predicted.

\begin{figure*}
\includegraphics[width=85mm,height=70mm,angle=0.0]{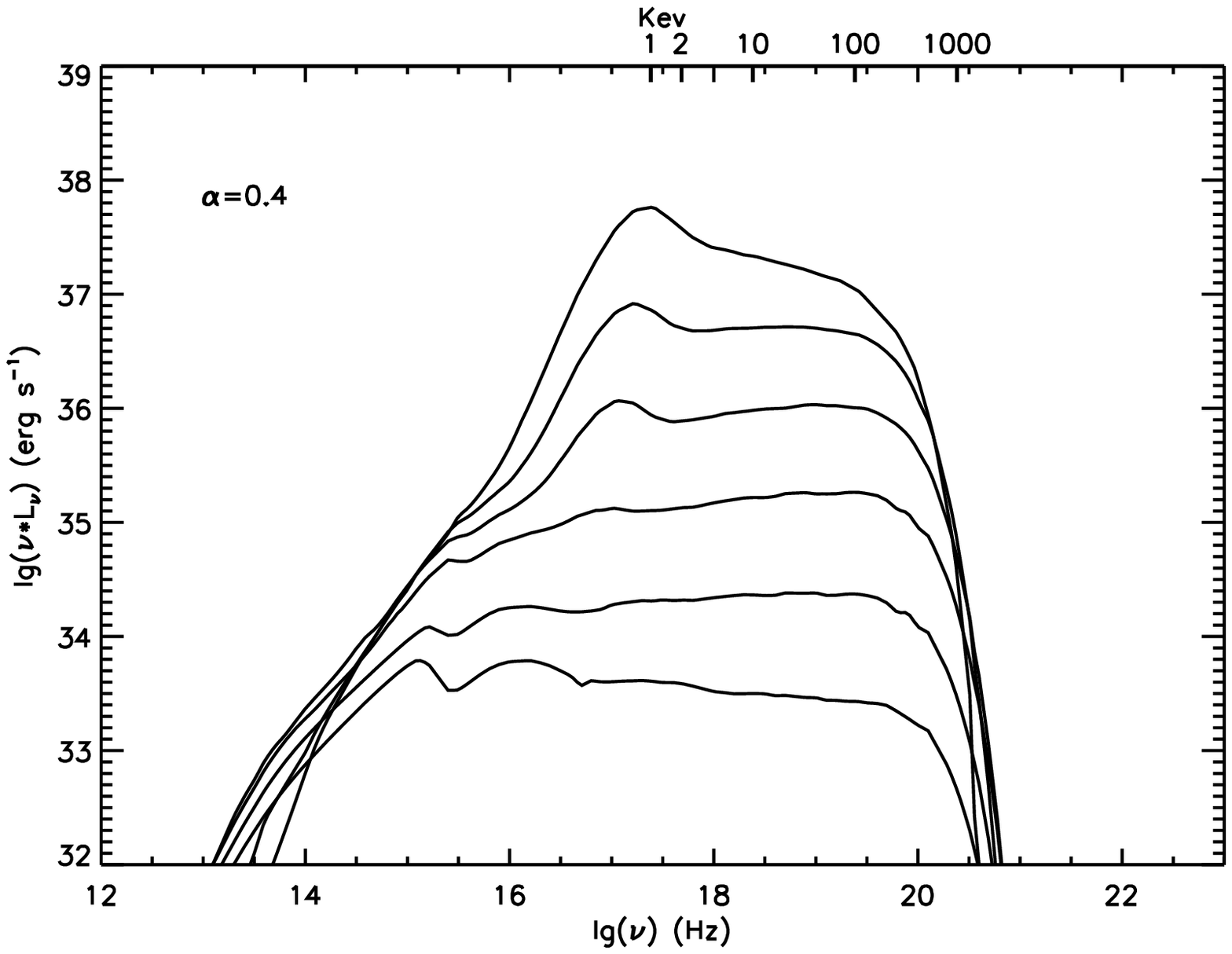}
\caption{\label{alpha04}  The emergent spectra for different mass
accretion rate. In the calculation, $m=10$, $\alpha=0.4$,
$\beta=0.8$ and $a=0.15$ are adopted. From the bottom up, the mass
accretion rates are $\dot m=0.005$, $0.01$, $0.02$, $0.03$, $0.05$,
$0.1$ respectively.}
\end{figure*}

To check the   effect of viscosity parameter $\alpha$, we plot
the emergent spectra for $\alpha=0.4$, $\dot m=$ $0.1$, $0.05$,
$0.03$, $0.02$, $0.01$, $0.005$ respectively in Figure
\ref{alpha04}. With decrease of mass accretion rate    from
0.1 to 0.02, the hard X-ray photon index $\Gamma_{\rm 2-10 keV}$
decreases from   2.32 to 1.86, then increases up to $2.13$
when $\dot m$ is down to $0.005$. The evolution of  $\Gamma_{\rm
2-10 keV}$ with Eddington ratio $\lambda$ is plotted in Figure
\ref{L-gama} with   red line. It's   clear that the
V-shaped relation between $\Gamma$ and $\lambda$ also holds
for $\alpha=0.4$.

We point out that in our calculations the mass accretion rate $\dot
m$ is assumed to be all in the corona at the condensation radius
$r_{\rm d}$. This is reasonable at low accretion rates because the
disk gas is completely evaporated into corona outside $r_{\rm d}$.
At accretion rate higher than the maximal evaporation rate
($\backsim 0.03$), this assumption is not valid any more. Only when
there is hot gas accreted from outer boundary, can the corona
accrete at an accretion rate larger than the maximal evaporation
rate. We note that, with a higher viscosity parameter ($\alpha
> 0.3$), the maximal accretion rates can be raised as high as 0.1 (Qiao
\& Liu 2009). Nevertheless, the positive correlation between
$\Gamma$ and $\lambda$ from $\dot m \backsim 0.01$ towards high
accretion rate does not change, as seen from Figure \ref{L-gama}.

\section{Comparison with observations}

H1743-322 is an X-ray transient, which was first discovered with the
Ariel 5 (Kaluzienski \& Holt 1977) and HEAO 1 (Doxsey et al. 1977)
satellites in 1977 August and then by the X-ray observatories RXTE
and INTEGRAl in the outburst 2003. The X-ray observations reveal
that the source   was going through different spectral   states
before fading at the end of 2003 (Markwardt \& Swank 2003; Homan et
al. 2003; Kretschmar et al. 2003; Grebenev et al. 2003; Tomsick \&
Kalemci 2003; Parmar et al. 2003; Joinet et al. 2005). Kalemci et
al.(2006)   investigated the decay of H1743-322 , i.e., the
transition from thermal dominated state to low/hard state in 2003,
and   studied the evolution of the spectra  in detail.   Up to date,
a wealth of observations indicate that the system is   an accreting
black hole with a central black hole mass of $\sim 10 M_{\odot}$
(Steiner et al. 2012), although the measurement of the central black
hole mass is still uncertain. Using the observed proper motions of
the X-ray jets, Corbel et al. (2005)   inferred an upper limit to
the source distance of $10.\pm 2.9 \rm kpc$. Steiner et al. (2012)
derived a distance of $8.5\pm 0.8 \rm kpc$ by applying a simple and
symmetric kinematic model to the trajectories of the two-sided jets.

\begin{figure*}
\includegraphics[width=85mm,height=70mm,angle=0.0]{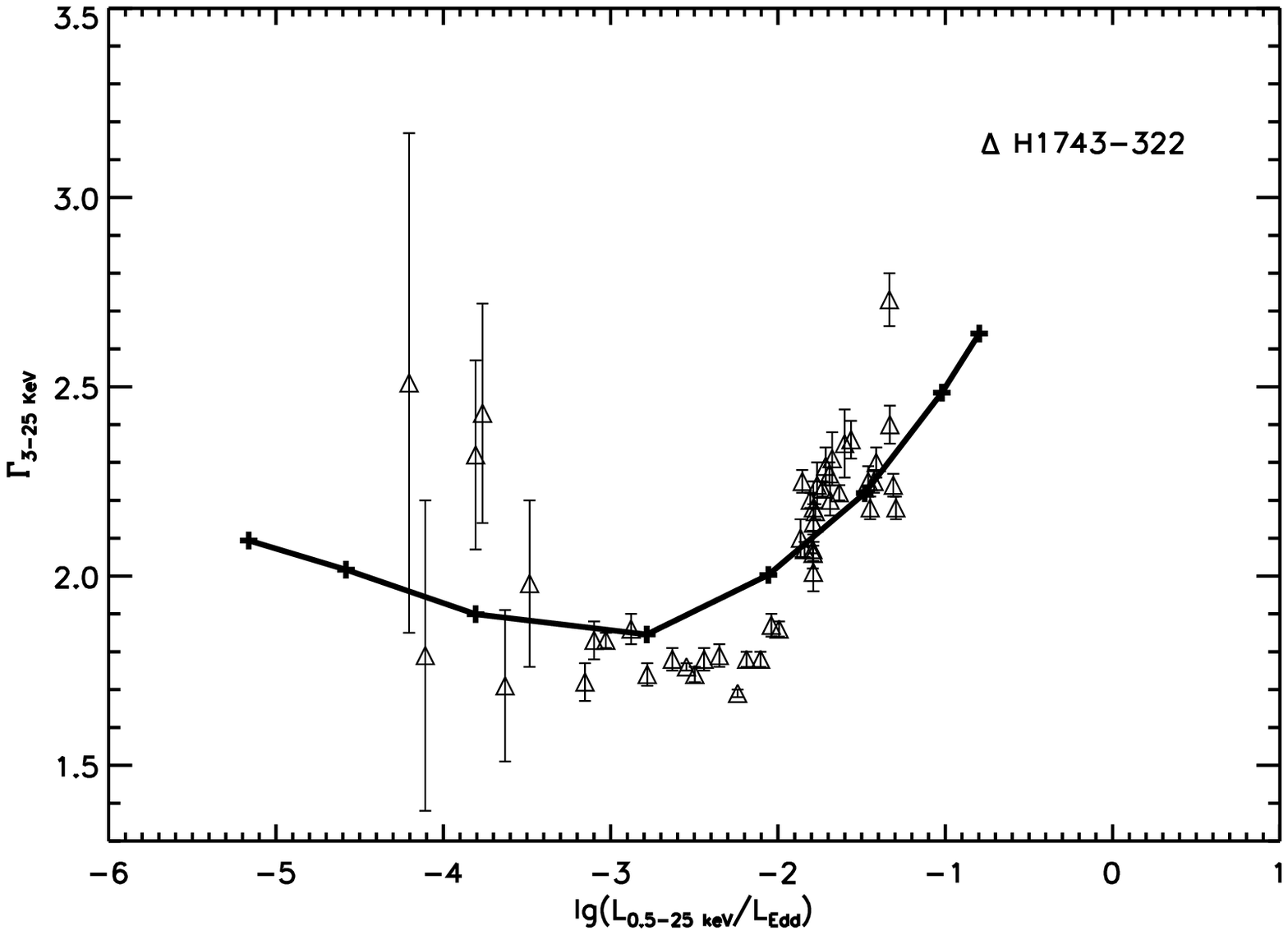}
\caption{\label{index}} Evolution of hard X-ray photon index
$\Gamma_{\rm 3-25keV}$ with $\rm lg(L_{0.5-25 \rm keV}/L_{\rm
Edd})$. The sign $\bigtriangleup$ is the observed data of H1743-322
during the decay in 2003, $m=10$ and $d=11 \rm kpc$ are adopted. The
solid line is the theoretical relation by taking $m=10$,
$\alpha=0.3$, $\beta=0.8$, and $a=0.15$.
\end{figure*}

Based on the observations of  Kalemci et al. (2006), Wu et al.
(2008) plotted the relation between $\Gamma_{\rm 3-25 keV}$ and
$L_{\rm 0.5-25 keV}/L_{\rm Edd}$ during the decay of H1743-322 in
2003 by taking the central black hole $m=10$ and distance $d=11 \rm
kpc$ respectively (see the black $\bigtriangleup$ style in Figure
\ref{index}). With $m=10$, $\alpha=0.3$, and $\beta=0.8$, the
theoretical spectra are calculated and shown in figure \ref{sp} for
different accretion rates,  $\dot m=0.003$, $0.005$, $0.01$, $0.02$,
$0.03$, $0.05$, $0.08$, $0.1$. For comparison with observations, the
model predicted relation between $\Gamma_{\rm 3-25 keV}$ and $L_{\rm
0.5-25 keV}/L_{\rm Edd}$ calculated from the spectra is plotted in
Figure \ref{index}. It can be seen that the model prediction is
roughly in agreement with the observations for reasonable
assumptions about the black hole mass and the distance to the
source.

Nevertheless,  the observational  $\Gamma$ shows a steeper hardening
at low Eddington ratios and then a sharper transition at high
Eddington ratios than the theoretical prediction, as shown in Figure
\ref{index}. This could be a combined effect of approximations and
chosen parameters involved  in the calculations. For example, the
self-similar solution used in this work deviates from the global
solution; The strength of magnetic field, the value of albedo and
the direct heating to corona electrons could  all depend on the
accretion rates, thereby affect the slope of $\Gamma$ shown in
Figure \ref{index}.  Some additional physics is needed to precisely
determine these parameters, which is beyond the present study.

\section{Conclusion}
We investigate the dependence of hard photon index $\Gamma$ on
Eddington ratio $\lambda$ within the framework of disk
evaporation/condensation model. In the paper, we update the
calculation of Meyer et al. (2006), Liu et al. (2007), Taam et al.
(2008) and Qiao \& Liu (2012), and self-consistently consider
  the cooling mechanisms including Compton scattering of the
soft photons from the underlying accretion disk, and the
bremsstrahlung, synchrotron and  self-Compton radiations to
calculate the structure of the disk and corona  and the spectrum
from the disk and corona. Our model can reproduce the positive
correlation between $\lambda$ and $\Gamma$ for $\lambda \gtrsim
0.01$, and the anti-correlation between $\lambda$ and $\Gamma$ for
$\lambda \lesssim 0.01$ observed in black hole X-ray binaries. We
show, as an example,  that the model can roughly explain the
observed evolution of $\Gamma_{\rm 3-25keV}$ with $L_{\rm 0.5-25
keV}/L_{\rm Edd}$ for H1743-322 in the outburst and decay of 2003
from thermal dominated state to low/hard state.

We appreciate the referee for his/her very help suggestions and
comments. We thank Qing-wen Wu for providing the observational data
of black hole X-ray binaries. We also thank R. E. Taam and B. Czerny
for comments and suggestions. ELQ appreciates the very useful
discussions with Xin-lin Zhou. This work is supported by the the
National Natural Science Foundation of China (grant 11033007 and
11173029), by the National Basic Research Program of China-973
Program 2009CB824800 and by the foundation for young researcher at
National Astronomical Observatories, Chinese Academy of Sciences.

%=============================================================================
%=============================================================================

%----------------------------------------------------------------------

%-----compare with observations---------------------------------------

\begin{table*}\label{T:result}
\begin{center}
\begin{minipage}{\linewidth}
\caption{Condensation and spectral features of the inner disk and
corona around the black hole of 10 $M_{\odot}$ }
\begin{tabular}{cc|ccccccc}
\hline\hline
&&&&&&&& \\
$\alpha$ & $\dot m$  & $r_{\rm d}$ & $\dot m_{\rm cnd}$ & $\dot m
_{\rm cnd}$/ $\dot m$   &$L/L_{\rm Edd}$ &
$T_{\rm eff,max} (\rm keV)$     & $\Gamma_{\rm 2-10 keV}$ \\
&&&&&&&& \\
\hline
0.3       & 0.1     & 400  &  $8.24\times10^{-2}$       &82\%&  $1.18\times10^{-1}$   &0.37  & 2.82\\

0.3       & 0.08    & 294  &  $6.16\times10^{-2}$       &77\%&  $9.66\times10^{-2}$   &0.35  & 2.63\\

0.3       & 0.05    & 162  &  $3.10\times10^{-2}$       &62\%&  $4.95\times10^{-2}$   &0.31  & 2.28\\

0.3       & 0.03    & 80   &  $1.22\times10^{-2}$       &41\%&  $1.60\times10^{-2}$   &0.25  & 2.01\\

0.3       & 0.02    & 28   &  $3.51\times10^{-3}$       &12\%&  $3.92\times10^{-3}$   &0.21  & 1.81\\

0.3       & 0.01    &      &                            &      &  $7.43\times10^{-4}$  &      & 1.85\\
0.3       & 0.005   &      &                            &      &  $1.39\times10^{-4}$  &      & 2.07\\
0.3       & 0.003   &      &                            &      &  $4.31\times10^{-5}$  &      & 2.20\\

\hline
0.4       & 0.1     & 266  &  $6.50\times10^{-2}$      &65\% &    $1.48\times10^{-1}$  &0.38    & 2.32\\
0.4       & 0.05    & 106  &  $1.88\times10^{-2}$      &38\% &    $3.16\times10^{-2}$  &0.29    & 1.98\\
0.4       & 0.03    & 42   &  $5.23\times10^{-3}$      &17\% &    $6.22\times10^{-3}$  &0.22    & 1.86\\
0.4       & 0.02    & 14  &   $9.56\times10^{-4}$      &5\%  &    $1.16\times10^{-3}$  &0.15    & 1.86\\
0.4       & 0.01    &      &  $1.52\times10^{-1}$      &       &  $3.75\times10^{-4}$  &        & 1.92\\
0.4       & 0.005   &      &  $1.52\times10^{-1}$      &       &  $7.35\times10^{-5}$  &        & 2.13\\
\hline \hline
\end{tabular}
\\
\\
Note.--- With black hole mass $m$=$10$, $\alpha=0.3$ and $0.4$, for
different mass accretion rate $\dot m$, the size of the inner disk
$r_{\rm d}$, condensation rate $\dot m_{\rm cnd}$ integrated from
the condensation radius to $3 R_{\rm S}$, the ratio of condensation
rate to the total accretion rate $\dot m_{\rm cnd} / \dot m$, the
Eddington ratio $L/L_{\rm Edd}$, the maximum temperature of the
inner disk $T_{\rm eff,max}$ and the hard X-ray photon index
$\Gamma_{\rm 2-10 keV}$ are listed. All the calculation, $m=10$,
$\beta=0.8$ and $a=0.15$ are adopted.
\end{minipage}
\end{center}
\end{table*}

\end{document}